\documentstyle{article}

\title{
Quantum solitons of the nonlinear $\sigma$-model with broken
chiral symmetry}

\author{
A.~Kostyuk and A.~Kobushkin \\
Bogolyubov Institute for theoretical physics, \\
Ukrainian Academy of Sciences,\\
Kiev 143, Ukraine \\
\and N.~Chepilko
\\Institute of Physics,\\
Ukrainian Academy of Sciences, \\
Kiev 028, Ukraine \\
\and T.Okazaki \\
Physics Laboratory, Sapporo Campus,\\
Hokkaido University of Education,\\
Sapporo 002, Japan
}

\date{July 28, 1993}

\begin{document}

\maketitle
\begin{abstract}
It is proved that the quantum-mechanical consideration of global
breathing of a hedgehog-like field configuration leads
to the dynamically stable soliton solutions in the nonlinear
$\sigma$-model without the Skyrme term. Such solutions exist only
when chiral symmetry of the model is broken.
\end{abstract}

\section{Introduction.}

It is well known that due to the Derric theorem there are no
topologically non-trivial classical soliton solutions in the nonlinear
$\sigma$-model when the Skyrme term is omitted.

However, recently it was assumed \cite{Quantum,Jain,Geom} that quantum
consideration of global breathing of hedgehog-like field
configuration in the nonlinear $\sigma$-model without the Skyrme term
leads to stable stationary soliton solutions. They are supposed to
vanish in the classical limit ($\hbar \rightarrow 0$) and in this
sense are called ``quantum solitons''.

Unfortunately, in the above papers the chiral angle was not determined
from the field equations. It either was approximated by the variation
method on some more or less suitable trial function family \cite{Geom},
or was postulated by some phenomenological reasons \cite{Jain,Jain1}.

In the present work the field equation for hedgehog-like quantum
globally breathing and rigidly rotating soliton in the nonlinear
$\sigma$-model with broken chiral symmetry and without the Skyrme
term is studied and numerical solution of this equation is
obtained\footnote{\protect\small Similar equation was also discussed
early \cite{Scale}, but now we do not include so-called quantum
corrections arising from quantum-mechanical treatment of collective
coordinates in the initial Lagrangian.}. We show that the model has
quantum soliton solutions, but only in the case when the chiral
symmetry is broken. So the numerical estimations done in
Refs.\cite{Jain,Geom,Jain1}, where the chiral symmetry limit was
assumed, are incorrect.

The paper is organized as follows.

Sec.\ref{Problem} contains a formulation of the nonlinear spectral
in\-te\-gro-dif\-fer\-en\-tial problem describing in terms of
collective coordinates a rotating and breathing soliton with
topological charge $1$. This problem consists of a Schr\"odinger
equation describing the dynamics of the collective coordinates and a
differential equation for the chiral angle with boundary conditions.

Sec.\ref{Properties} is devoted to the analysis of the
boundary-value problem for the chiral angle and its exact numerical
solutions.

In Sec.\ref{Solution} the Schr\"odinger equation spectrum is
constructed, the exact solution of the integro-differential problem
corresponding to lowest-lying breathing states with isotopic spin
$\displaystyle \frac{1}{2} $ and $\displaystyle \frac{3}{2} $ are
obtained. The latter are considered as models of lightest baryons.

An analysis of the results and discussions are given in
Sec.\ref{Conclusion}.

\section{The nonlinear integro-differential spectral problem for
the rigidly rotating and globally breathing soliton.}

\label{Problem}
The initial Lagrangian density of the nonlinear $\sigma$-model with
broken chiral symmetry is given by
\begin{equation}
{\cal L}=-\frac{f_{\pi}^{2}}{4} \mbox{\rm Tr}(J_{\mu}J^{\mu})
+\frac{1}{16}f_{\pi}^{2}m_{\pi}^{2} \mbox{\rm Tr}(\vec{\tau} U
\vec{\tau} U^{+}-3).
\label{l-ian}
\end{equation}
The following notations are used: $\displaystyle J_{\mu} \equiv U^{+}
\frac{\partial U} {\partial x^{\mu}} $, $U=U(t,\vec{x})$ is the
$2 \times 2$ chiral field matrix; $x^{\alpha}=(t, \vec{x})$,
$\alpha=0,1,2,3$
are the coordinates in the Minkowski space with the signature
$(+ - - -)$, $\vec{\tau}=(\tau_{1},\tau_{2},\tau_{3})$ are the
isotopic Pauli matrices, $f_{\pi}=93 MeV$ stands for the
pion decay constant, $m_{\pi}$ is a parameter of the theory and,
as it will be discussed later, need not to be equal to physical
pion mass.

Instead of the term breaking chiral symmetry commonly used
\begin{equation}
\frac{1}{4}f_{\pi}^{2}m_{\pi}^{2} \mbox{\rm Tr}(U+U^{+}-2),
\label{term1}
\end{equation}
we choose the term \cite{Scale}
\begin{equation}
\frac{1}{16}f_{\pi}^{2}m_{\pi}^{2} \mbox{\rm Tr}(\vec{\tau} U
\vec{\tau} U^{+}-3),
\label{term2}
\end{equation}
which considerably simplifies the further analysis. It can be shown
that the term (\ref{term2}) corresponds to chiral symmetry breaking
term (2.10) of Ref.\cite{Marleau} with $k=2$. For the weak pion
fields (which take place, e.g., at the large distances from the
soliton center) both terms (\ref{term1}) and (\ref{term2}) are
equivalent.

Here we shall consider the solution in the form of the rigidly
rotating and globally breathing hedgehog ansatz used in Refs.
\cite{Jain,Scale}:

\begin{equation}
U=A(t) \exp \left[ i \vec{\tau} \hat{x} \theta (z) \right] A^{+}(t),
\label{h-hog}
\end{equation}
where $\displaystyle \hat x =\frac{\vec{x}}{|\vec{x}|} $,
$A \in SU(2),$ $A=a_{0}(t)+i\vec{\tau}\vec{a}(t),$
$a_{0}^{2}+\vec{a}^{2}=1,$ $r=|\vec{x}|,$
\begin{equation}
z=\frac{r}{\lambda(t)}.
\label{z}
\end{equation}
The chiral angle $\theta(z)$  satisfies the boundary conditions:
\begin{equation}
\begin{array}{cc}
\left. \theta(z) \right|_{z=0} =\pi, &
\left. \theta(z) \right|_{z \rightarrow \infty}=0,
\end{array}
\label{boundary}
\end{equation}
which correspond to the field configuration with the topological
charge $1$. The quantities $a_{p}=(a_{0},\vec{a})$,
$p=0,1,2,3$ and the homogeneous scale transformation parameter
$\lambda(t)$ are the quantum collective coordinates describing
rotation and breathing of the soliton, respectively.

In terms of these collective coordinates the dynamical system
Lagrangian

\begin{equation}
\Lambda = \int d^{3}x {\cal L}
\end{equation}
is written as:
\begin{equation} \Lambda =
\frac{1}{2} m[\theta] \dot{q}^{2}-v(q)+ \frac{1}{2} \mu [\theta]
q^{2} \vec{W}^{2}
\label{l-func}
\end{equation}
\begin{equation}
v(q)= \eta[\theta] q^{2/3} +
\frac{3}{4}m_{\pi}^{2} \mu [\theta] q^{2},
\label{p-ial}
\end{equation}
where $q=\lambda^{3/2},$
$\vec{W}=2(a_{0}\dot{\vec{a}}-\vec{a}\dot{a}_{0}+
\vec{a}\times\dot{\vec{a}})$
is the rotation velocity. The quantities
\begin{equation}
m[\theta]=\frac{16}{9} \pi f_{\pi}^{2}  \int_{0}^{\infty}
dz z^4 \left(\frac{d\theta}{dz}\right)^2,
\label{m}
\end{equation}
\begin{equation}
\mu[\theta]=\frac{8}{3} \pi f_{\pi}^{2}  \int_{0}^{\infty}
dz z^2 \sin ^2 \theta
\label{mu}
\end{equation}
and
\begin{equation}
\eta[\theta]=2 \pi f_{\pi}^{2}  \int_{0}^{\infty}
dz \left(z^2 \left(\frac{d\theta}{dz}\right)^2
+ 2 \sin ^2 \theta \right)
\label{eta}
\end{equation}
are functionals on the chiral angle $\theta(z)$.

Introducing the new generalized coordinates $q_{p}=qa_{p}$, one
obtains \cite{Jain,Scale} the Hamiltonian operator, corresponding
to the Lagrangian (\ref{l-func})
\begin{equation}
H=-\frac{1}{2m} \frac{1}{q^3} \frac{\partial}{\partial q}
\left( q^3 \frac{\partial}{\partial q} \right)
+ \frac{\vec{J}^2}{2 \mu q^2} + v(q)
\label{h-ian}
\end{equation}
( $\vec{J}$ stands for the angular momentum operator). In this case
the normalization condition for the wave function satisfying the
Schr\"odinger equation
\begin{equation}
H \psi = E \psi
\label{schrod}
\end{equation}
is
\begin{equation}
\int d\Omega \int_{0}^{\infty} dq q^3 |\psi|^2 = 1 .
\label{norm}
\end{equation}
(the first integral denotes integration over the angular variables
in the space of generalized coordinates $q_{p}$ ).
The wave function $\psi(q)$ is supposed to satisfy the boundary
conditions
\begin{equation}
\left.\psi \right|_{q=0}
= \left.\psi \right|_{q \rightarrow \infty} = 0.
\label{bounpsi}
\end{equation}
\sloppy
Separating in Eq.(\ref{schrod}) the rotational and the breathing
variables $\psi=\psi_{nj}(q)\psi_{j}(a_{p})$ one gets the
Schr\"odinger equation for the new radial wave function
$\varphi_{nj}(q)=q^{3/2}\psi_{nj}(q)$ in the following form
\begin{equation}
-\frac{1}{2m} \frac{d^2}{dq^2} \varphi_{nj}(q)
+(v_{j}(q)-E_{nj}) \varphi_{nj}(q)=0 ,
\label{schrod1}
\end{equation}
\begin{equation}
v_{j}(q)=\frac{j(j+1)}{2\mu q^2} + \frac{3}{8m q^2} + \eta q^{2/3}
+\frac{3}{4} m_{\pi}^{2} \mu q^2
\label{p-ialj}
\end{equation}
with the normalization condition
\begin{equation}
\int_{0}^{\infty} dq |\varphi_{nj}(q)|^2=1 .
\label{norm1}
\end{equation}

$\displaystyle j=0,\frac{1}{2},1,\frac{3}{2},\ldots $ and
$n=0,1,2,\ldots$ are quantum numbers of isotopic and radial
excitation respectively.

When $m_{\pi} \not= 0,$ it is convenient to reduce
Eqs.(\ref{schrod1})--(\ref{norm1}) to the form
\begin{equation}
\frac{d^2}{d\chi^2} \tilde{\varphi}_{nj}(\chi)-
\left(\frac{\beta_{j}^2-1}{4\chi^2}+\chi^2+\gamma \chi^{2/3} -
\varepsilon_{nj} \right) \tilde{\varphi}_{nj}(\chi) = 0,
\label{schrod2}
\end{equation}
\begin{equation}
\int_{0}^{\infty} d\chi |\tilde{\varphi}_{nj}(\chi)|^2=1
\label{norm2}
\end{equation}
by the replacement
\begin{equation}
\begin{array}{ccc}
\displaystyle \chi=q\sqrt{m_{\pi}\sqrt{\frac{3}{2}m\mu}}, & &
\displaystyle \tilde{\varphi}_{nj}(\chi)
=\frac{\varphi_{nj}(q)}
{\left( m_{\pi}\sqrt{\frac{3}{2}m\mu} \right)^{1/4}}.
\end{array}
\label{r-ment}
\end{equation}
Here we use the following notation
\begin{eqnarray}
\beta_{j}^{2} &=& 4\left(\frac{m}{\mu}j(j+1)+1\right),
\label{beta} \\
\gamma &=& \frac{2}{m_{\pi}^{4/3}}\left( \frac{4}{9}\frac{m}{\mu}
\right)^{1/3} \frac{\eta}{\mu^{1/3}},
\label{gamma} \\
\varepsilon_{nj} &=& \frac{2E_{nj}}{m_{\pi}} \sqrt{\frac{2}{3}
\frac{m}{\mu}}.
\label{eps}
\end{eqnarray}

To solve Eq.(\ref{schrod2}) the quantities $\beta_{j}$ and $\gamma$
must be determined. They are expressed in terms of $m$,$\mu$ and
$\eta$ defined by (\ref{m})--(\ref{eta}), and thus are functionals
on the chiral angle $\theta(z)$.

It was shown \cite{Scale} that $\theta(z)$ satisfies the
following equation
\begin{equation}
\langle nj | \delta_{\theta} \Lambda | nj \rangle = 0
\label{principe}
\end{equation}
where $|nj\rangle$ is the eigenvector of the Hamiltonian operator
(\ref{h-ian}).
The Lagrangian (\ref{l-func}) depends on $\theta(z)$ through three
functionals $m$,$\mu$ and $\eta$, therefore one can rewrite
Eq.(\ref{principe}) in the following form

\begin{equation}
a\delta_{\theta}m + b\delta_{\theta}\mu + c\delta_{\theta}\eta =0.
\label{principe1}
\end{equation}

The quantities
\begin{eqnarray}
a &=& \langle nj |\frac{1}{2}\dot{q}^2| nj \rangle,
\label{a}\\
b &=& \langle nj |\frac{1}{2} q^2 \vec{W}^2 -
\frac{3}{4} m_{\pi}^2 q^2 | nj \rangle,
\label{b} \\
c &=& -\langle nj |q^{2/3}| nj \rangle
\label{c}
\end{eqnarray}
are functionals of the quantum system wave function.

Using the notations
\begin{eqnarray}
z_{0} &=& \sqrt{\frac{9}{8} \frac{|c|}{a}},
\label{z0} \\
\alpha &=& \frac{3}{4} \frac{b}{a}
\label{alpha}
\end{eqnarray}
and introducing the new variables
\begin{equation}
\theta(z) = \theta(z_{0}\xi) = F(\xi)
\label{repl}
\end{equation}
one obtains the differential equation for the chiral angle:
\begin{equation}
\left( \xi^{2}(1 - \xi^{2})F' \right)' -
\left( 1 - \alpha \xi^{2} \right) \sin 2F = 0 .
\label{diffeq}
\end{equation}
It should be supplemented by the boundary conditions

\begin{equation}
\left. F \right|_{\xi=0}=\pi ,
\label{bound0}
\end{equation}
\begin{equation}
\left. F \right|_{\xi \rightarrow \infty}= 0 .
\label{boundinf}
\end{equation}

Eqs.(\ref{diffeq}) and (\ref{schrod}) are related by the integral
expressions for their coefficients (\ref{m})--(\ref{eta}) and
(\ref{alpha}),(\ref{a}),(\ref{b}). Thus the rotating and breathing
soliton is described by the nonlinear integro-differential spectral
problem.

When the condition $m_{\pi} \not= 0$  is fulfilled, it is
convenient to modify the integro-differential problem replacing
the Eq. (\ref{schrod}) by (\ref{schrod2}) and to express the
functionals (\ref{a})--(\ref{c}) in terms of wave function
$\tilde{\varphi}(\chi)$:

\begin{eqnarray}
a &=& \frac{m_{\pi}}{2m}\sqrt{\frac{3}{2}\frac{\mu}{m}}
\int_{0}^{\infty} \tilde{\varphi}^{\ast}
\left( \frac{3}{4} \frac{1}{\chi^2} -
\frac{d^{2}}{d \chi^{2}} \right)
\tilde{\varphi} d \chi ,
\label{a1}  \\
b &=& \frac{m_{\pi}}{2m}\sqrt{\frac{3}{2}\frac{m}{\mu}}
\int_{0}^{\infty} \left( \frac{m}{\mu} \frac{j(j+1)}{\chi^2}
- \chi^{2} \right) |\tilde{\varphi}|^{2} d \chi ,
\label{b1} \\
c &=& - \frac{1}{\left(m_{\pi}\sqrt{\frac{3}{2}m \mu}\right)^{1/3}}
\int_{0}^{\infty} \chi^{2/3} |\tilde{\varphi}|^{2} d \chi .
\label{c1}
\end{eqnarray}
Applying the quantum virial theorem to the Schr\"odinger equation
(\ref{schrod2}) one has
\begin{equation}
\int_{0}^{\infty} \tilde{\varphi}^{\ast} \frac{d^{2}}{d \chi^{2}}
\tilde{\varphi} d \chi
 = \int_{0}^{\infty} \left( \frac{\beta_{j}^2-1}{4 \chi^2}
- \chi^{2} -\frac{1}{3}\gamma \chi^{2/3} \right)
|\tilde{\varphi}|^{2} d \chi .
\label{virial}
\end{equation}
Substituting the last expression into (\ref{a1}) and taking into
account (\ref{beta}) one gets
\begin{equation}
a  =  - \frac{m_{\pi}}{2m}\sqrt{\frac{3}{2}\frac{\mu}{m}}
\int_{0}^{\infty} \left( \frac{m}{\mu} \frac{j(j+1)}{\chi^2}
- \chi^{2} - \frac{1}{3} \gamma \chi^{2/3} \right)
|\tilde{\varphi}|^{2} d \chi .
\label{a2}
\end{equation}

\section{Properties of the chiral angle.}

\label{Properties}
Let us investigate the behavior of the solution of
Eq.(\ref{diffeq}) with different $\alpha$. This equation has three
singular points $\xi=0,1,\infty$.

Taking into account (\ref{bound0}) one obtains the following solution
for $\xi \ll 1$ region
\begin{equation}
F = \pi + \xi \left. F' \right|_{\xi=0} + O(\xi^3).
\label{regular0}
\end{equation}
On the other hand, the chiral angle will be regular at $\xi=1$ only
when the following condition is fulfilled
\begin{equation}
\left. F' \right|_{\xi=1} +
\frac{1}{2}(1-\alpha) \left. \sin 2F \right|_{\xi=1}=0 .
\label{regular1}
\end{equation}
One has to make such a choice of the quantities
$\left. F' \right|_{\xi=0}$ and $\left. F \right|_{\xi=1}$ that the
two solutions of Eq.(\ref{diffeq}) starting from points $\xi=0$ and
$\xi=1$ with boundary conditions (\ref{regular0}) and
(\ref{regular1}), respectively,  will be smoothly joint in arbitrary
point of the interval $(0,1)$ (See Appendix). The numerical analysis
shows that for every $\alpha$ there exist two sets of these values.
Although the existence of another set is not excluded completely,
we have not succeeded in obtaining it.

So on $[1,\infty)$ the boundary-value problem
(\ref{diffeq})--(\ref{boundinf}) is reduced to two initial-value
problems. Their initial conditions $\left. F' \right|_{\xi=1}$ and
$\left. F \right|_{\xi=1}$ and, consequently, their solutions are
uniquely determined by $\alpha$ value. So $\alpha$ is the spectral
parameter of boundary-value problem, because the condition
(\ref{boundinf}) can be satisfied only by a special choice
of $\alpha$.

The two obtained sets of values $\left. F' \right|_{\xi=0}$ and
$\left. F \right|_{\xi=1}$ are turned into each other by the
transformation $F \rightarrow 2\pi-F$. Eq.(\ref{diffeq}) is also
invariant under this transformation. So the two solutions
of the initial-value problems corresponding to these two sets
have the same symmetry.

One of them tends to the following values at large $\xi$
\begin{equation}
\left. F \right|_{\xi \rightarrow \infty} = \left\{
\begin{array}{ccr}
\displaystyle \frac{\pi}{2} & {\rm if} & \alpha > 0 \\[1.5ex]
\displaystyle \sim 0.86     & {\rm if} & \alpha = 0 \\[1.5ex]
\displaystyle    0          & {\rm if} & \alpha < 0 .
\end{array} \right.
\label{liminf}
\end{equation}
Using the transformation $F \rightarrow 2\pi-F$ one gets that
the other solution does not satisfy (\ref{boundinf}) for
any $\alpha$. (For $\alpha < 0$ it leads to the solution
with topological charge $-1$.)
So the boundary-value problem (\ref{diffeq})--(\ref{boundinf}) has
solution under the condition
\begin{equation}
\alpha < 0
\label{condalph}
\end{equation}
only.
It is equivalent to
\begin{equation}
m_{\pi}^{2} > \frac{2}{3}
\frac{\langle nj |\frac{1}{2} q^2 \vec{W}^2 | nj \rangle}
{\langle nj | q^2 | nj \rangle}.
\label{condmpi}
\end{equation}
The last expression has been obtained by comparing the expression
(\ref{condalph}) with (\ref{a}),(\ref{b}) and (\ref{alpha}).

Note that right hand side of (\ref{condmpi}) is
always positive, so the boundary-value problem
(\ref{diffeq})--(\ref{boundinf}) has no solution in the chiral
approximation ($m_{\pi}=0$).

It should be noted here that some authors \cite{Kobayashi,Asano},
using variational method, have also concluded that the globally
breathing and rigidly rotating hedgehog-like quantum soliton does
not exist in the chiral limit  of the nonlinear $\sigma$-model
without the Skyrme term.

Provided that the condition (\ref{condalph}) is fulfilled,
Eq.(\ref{diffeq}) is asymptotically reduced to
\begin{equation}
\xi^{2}F'' + 4 \xi F' - 2 \alpha F = 0
\label{asympteq}
\end{equation}
at $\xi \gg 1$.
The solution of this equation, coinciding with the solution of
boundary-value problem (\ref{diffeq})--(\ref{boundinf}) for
$\xi \gg 1$, can be expressed in terms of elementary functions
\begin{equation}
F(\xi) = \left\{
\begin{array}{lcr}
\displaystyle C_{1} \xi^{\kappa_{1}} +  C_{2} \xi^{\kappa_{2}},
&\displaystyle {\rm if} &\displaystyle -\frac{9}{8}<\alpha<0 \\[1.5ex]
\displaystyle C_{1} \xi^{-3/2} \left( \ln \xi +  C_{2} \right),
&\displaystyle {\rm if} &\displaystyle \alpha=-\frac{9}{8} \\[1.5ex]
\displaystyle C_{1} \xi^{-3/2} \sin \left(\kappa_{0}\ln \xi + C_{2}
\right),
&\displaystyle {\rm if} &\displaystyle \alpha<-\frac{9}{8}
\end{array} \right.
\label{asymptsol}
\end{equation}
where
\begin{eqnarray}
\kappa_{1,2} &=& -\frac{3}{2} \pm
\sqrt{\left(\frac{3}{2}\right)^{2}+2\alpha} ,
\label{kappa12} \\
 \kappa_{0} &=& \sqrt{2|\alpha|-\left(\frac{3}{2}\right)^{2}} ,
\label{kappa0}
\end{eqnarray}
$C_{1}$ and $C_{2}$ are arbitrary constants.
It was shown numerically that for any $\alpha$ from the
interval $(-9/8,0)$  $C_{1}$ is nonequal to zero. So, it is easy to
mention that for all $\alpha<0$ the functionals (\ref{m}) and
(\ref{mu}) are divergent, because the chiral angle decrease too
slowly at infinity. In spite of that, the soliton
parameters are finite. Actually, according to the formulas
(\ref{beta})--(\ref{eps}) Eq. (\ref{schrod2}) contains
$m$, $\mu$ and $\eta$ through the ratios $\displaystyle
\frac{m}{\mu} $ and $\displaystyle \frac{\eta}{\mu^{1/3}} $ only.
Considering the ratios of the divergent integrals in the sense of
principal value, for instance:

\begin{equation}
\frac{m}{\mu}= \lim_{Z \rightarrow \infty}
\frac{\displaystyle \frac{16}{9} \pi f_{\pi}^2 \int_{0}^{Z} dz z^{4}
\left(\frac{d\theta}{dz}\right)^{2}}
{\displaystyle \frac{8}{3} \pi f_{\pi}^2 \int_{0}^{Z} dz z^{2}
\sin^{2} \theta},
\label{vp}
\end{equation}
and taking into account (\ref{asymptsol}),(\ref{kappa12}) and
(\ref{kappa0}) one gets finite numbers
\begin{equation}
\frac{m}{\mu}= \left\{
\begin{array}{lcr}
\displaystyle \frac{2}{3} \kappa_{1}^{2}
& {\rm if} &\displaystyle  -\frac{9}{8}<\alpha<0 \\[1.5ex]
\displaystyle \frac{4}{3} |\alpha|
& {\rm if} &\displaystyle \alpha \le -\frac{9}{8}
\end{array} \right.
\label{rat1}
\end{equation}
and
\begin{equation}
\begin{array}{ccc}
\displaystyle \frac{\eta}{\mu^{1/3}}=0
&\displaystyle {\rm for\ all} &\displaystyle \alpha<0.
\end{array}
\label{rat2}
\end{equation}

\section{\sloppy The exact soliton solutions for lowest-lying
quantum states.}

\label{Solution}
Substituting (\ref{b1}) and (\ref{a2}) into (\ref{alpha}) and
taking into account (\ref{rat2}) and (\ref{gamma}) one has
\begin{equation}
\alpha =- \frac{3}{4} \frac{m}{\mu}.
\label{alpha1}
\end{equation}
Comparing the last expression with (\ref{rat1}) one obtains the
algebraic equation for $\alpha$. For $-9/8<\alpha<0$ this equation
is written as
\begin{equation}
\frac{3}{2}\sqrt{\left(\frac{3}{2}\right)^{2}+2\alpha}=
\left(\frac{3}{2}\right)^{2}+2\alpha
\label{alg1}
\end{equation}
and has no solution, but for $\alpha \le -9/8$ it can be reduced to
\begin{equation}
\alpha=-|\alpha|,
\label{alg2}
\end{equation}
i.e. it is fulfilled identically.
Hence the integro-differential problem formulated in
Sec.\ref{Problem} has a solution for any $\alpha$ from
$(-\infty,-9/8]$. Still, for further calculations we have chosen
$\alpha=-9/8$, because, as seen in (\ref{asymptsol}), for
$\alpha<-9/8$ the chiral angle becomes oscillating at large distances
and the physical meaning of such solutions is unclear.

Taking into account (\ref{rat2}) and (\ref{gamma}) the
Eq.(\ref{schrod2}) can be reduced to
\begin{equation}
\frac{d^2}{d\chi^2} \tilde{\varphi}_{nj}(\chi)-
\left(\frac{\beta_{j}^2-1}{4\chi^2}+\chi^2 -
\varepsilon_{nj} \right) \tilde{\varphi}_{nj}(\chi) = 0.
\label{schrod3}
\end{equation}
The eigenvalue spectrum of this equation is as follows:
\begin{equation}
\varepsilon_{nj} = 2(2n+1)+\beta_{j},\ \ \ n=0,1,2,\ldots
\label{spectr}
\end{equation}
The solution of Eq.(\ref{schrod3}) corresponding to the ground
breathing state has the form
\begin{equation}
\tilde{\varphi}_{0j}(\chi)=
\frac{2}{\sqrt{\beta_{j}\Gamma(\beta_{j} /2)}}
\chi^{\frac{1}{2}(\beta_{j}+1)} \exp(-\frac{\chi^2}{2})
\label{wavef0}
\end{equation}
where $\Gamma(y)$ is a Gamma function.

Comparing the formulas (\ref{eps}),(\ref{spectr}) and (\ref{beta})
one gets the energy spectrum expressed in terms of $\alpha$:
\begin{eqnarray}
E_{nj} & = & \left( n+\frac{1}{2}+\frac{\beta_{j}}{4} \right)
\frac{3m_{\pi}}{\sqrt{2|\alpha|}},
\label{energy0} \\
\beta_{j} & = & 2\sqrt{\frac{4}{3}|\alpha|j(j+1)+1}.
\label{beta1}
\end{eqnarray}

It was shown in Ref.\cite{Geom} that the mean square radius of the
breathing soliton can be evaluated by
\begin{eqnarray}
\langle r^{2} \rangle_{nj} &=& \int_{0}^{\infty} d\chi
|\tilde{\varphi}_{nj}(\chi)|^2 \nonumber \\
& & \times\int_{0}^{\infty} dr r^{2} \left( -\frac{2}{\pi}
\sin^2 \left( \theta \left(\frac{\displaystyle r}
{\displaystyle \lambda(\chi)}\right)\right) \frac{d \theta
\left(\frac{r}{\lambda(\chi)}\right)}{dr} \right)
\label{radius}
\end{eqnarray}
Using (\ref{z}),(\ref{repl}) and (\ref{wavef0})  one obtains after
straightforward calculations:
\begin{equation}
\langle r^{2} \rangle_{0j} = \frac{2|\alpha|}{m_{\pi}^2}
\frac{\Gamma\left(\frac{\beta_{j}}{2}+\frac{1}{3}\right)
\Gamma\left(\frac{\beta_{j}}{2}+\frac{2}{3}\right)}
{\left(\Gamma\left(\frac{\beta_{j}}{2}\right)\right)^{2}}
\frac{\left(\beta_{j}+\frac{2}{3}\right)
\left(\beta_{j}+\frac{4}{3}\right)}
{\beta_{j}\left(\beta_{j}+2\right)}
\langle \xi^{2} \rangle ,
\label{radius0j}
\end{equation}
where
\begin{equation}
\langle \xi^{2} \rangle = -\frac{2}{\pi}
\int_{0}^{\infty} d \xi \xi^2 \sin^2 F(\xi)
\frac{d F(\xi)}{d \xi} .
\label{xi2}
\end{equation}
It should be mentioned that the soliton parameters $E_{nj}$ and
$\langle r^{2} \rangle_{0j}$ depend on the parameter $m_{\pi}$
only, and do not on the pion decay constant $f_{\pi}$.

Considering the solutions for the three lowest-lying breathing states
with the isotopic spin numbers $\displaystyle \frac{1}{2} $ and
$\displaystyle \frac{3}{2} $ as models of the particles $N(940)$,
$N(1440)$, $N(1710)$, $\Delta(1232)$, $\Delta(1600)$ and
$\Delta(1920)$, respectively, we have calculated some relations
between their static properties (the masses and the mean squared
 radius of the nucleon), which are independent on $m_{\pi}$ value.
The results of the calculations and the corresponding experimental
values are shown in the Table 1.
\begin{table}
\begin{center}
\caption{} The relations between some soliton static properties \\
being compared with corresponding experimental values. \\
\mbox{} \\
\begin{tabular}{||c|c|c||}  \hline\hline
\rule[-2.5ex]{0ex}{5.5ex} \makebox[5em]{} &
\makebox[8em]{Present model} & \makebox[8em]{Experiment} \\ \hline
\rule[-3.3ex]{0ex}{7.5ex}
$\displaystyle E_{0\frac{1}{2}} \langle r^{2}
\rangle_{0\frac{1}{2}}^{1/2} $
& $3.82$ & $3.38$ \\ \hline
\rule[-3.3ex]{0ex}{7.5ex}
$\displaystyle \frac{E_{1\frac{1}{2}}}{E_{0\frac{1}{2}}} $
& $1.83$ & $1.53$ \\ \hline
\rule[-3.3ex]{0ex}{7.5ex}
$\displaystyle \frac{E_{2\frac{1}{2}}}{E_{0\frac{1}{2}}} $
& $2.66$ & $1.82$ \\ \hline
\rule[-3.3ex]{0ex}{7.5ex}
$\displaystyle \frac{E_{0\frac{3}{2}}}{E_{0\frac{1}{2}}} $
& $1.43$ & $1.31$ \\ \hline
\rule[-3.3ex]{0ex}{7.5ex}
$\displaystyle \frac{E_{1\frac{3}{2}}}{E_{0\frac{1}{2}}} $
& $2.26$ & $1.70$ \\ \hline
\rule[-3.3ex]{0ex}{7.5ex}
$\displaystyle \frac{E_{2\frac{3}{2}}}{E_{0\frac{1}{2}}} $
& $3.09$ & $2.04$ \\ \hline\hline
\end{tabular}
\end{center}
\end{table}

\section{Conclusions and discussion.}

\label{Conclusion}
In the present paper we have proved that the quantum-mechanical
consideration of the global breathing of hedgehog-like field
configuration leads to the dynamically stable soliton solutions of
the nonlinear $\sigma$-model. Such solutions exist only when chiral
symmetry is broken.

The relations between some static properties of the solitons in ground
breathing state with isotopic spin $\displaystyle \frac{1}{2} $ and
$\displaystyle \frac{3}{2} $ appear to be close with accuracy 10--15\%
to those of nucleon and delta, respectively. However for the excited
states the predictions of the model are somewhat worse.

Nevertheless, the proposed model has some difficulties.

Firstly, the soliton masses and the mean squared radius depend on the
parameter $m_{\pi}$ and in order to reproduce the experimental values
one has to use $m_{\pi} \approx  2.7 \times \mbox{(the physical
pion mass)}$. We believe that this problem can be solved in the
framework of the quantum soliton model for the pion \cite{Pion}.
In such approach the physical pion mass arises as an eigenvalue of
corresponding Hamiltonian operator, but not as a parameter in the
initial Lagrangian.

Secondly, the slow asymptotic falloff of the profile function on the
large distances from the soliton center ($F(\xi) \sim \xi^{-3/2}
\ln \xi$) contradicts the Yukawa law. This means that the assumption
about homogeneous global breathing and rigid rotation is too rough and
does not describe the soliton external part correctly. We hope, this
problem may be solved by localization of the collective coordinate
proposed in Refs.\cite{Geom,Jadern}.

Thirdly, the physical meaning of oscillating solutions (when
$\displaystyle \alpha<-\frac{9}{8}$) is unclear.

\section*{Acknowledgment.}

We thank Prof.~K.~Fujii for helpful discussions.

\section*{\sloppy Appendix: The Numerical Calculation Scheme.}

Within the interval $[0,1]$ the chiral angle $F$ can be obtained
as a solution of the boundary-value problem consisting of the
differential equation (\ref{diffeq}) and boundary conditions
(\ref{bound0}) and (\ref{regular1}).

Because $\xi=0$ and $\xi=1$ are singular points of
Eq.(\ref{diffeq}), the conventional shooting method can not be used
in the present case. So it was modified as follows.

For any value $F'_{0}$ one obtains an initial-value problem with
initial conditions
\begin{equation}
\left. \frac{dF}{d\xi} \right|_{\xi=0}=F'_{0}
\label{abound0}
\end{equation}
and (\ref{bound0}). For  arbitrarily chosen point $\xi_{p} \in (0,1)$
this problem can be solved on the interval $[0,\xi_{p}]$ by, for
instance, the Runge-Kutta method.

Similarly, for any $F_{1}$ the problem with initial conditions
\begin{eqnarray}
\left. F \right|_{\xi=1} &=& F_{1} \label{abound1a}\\
\left. \frac{dF}{d\xi} \right|_{\xi=1} &=&\displaystyle -
\frac{1}{2}(1-\alpha) \sin 2F_{1}
\label{abound1b}
\end{eqnarray}
can be solved from point $\xi=1$ to $\xi=\xi_{p}$.

Let us introduce the following notations
\begin{eqnarray}
\Delta (F'_{0},F_{1}) &=& \left. F^{(0)} \right|_{\xi=\xi_{p}} -
\left. F^{(1)} \right|_{\xi=\xi_{p}}
\label{adelta1} \\
\Delta' (F'_{0},F_{1}) &=&\displaystyle \left. \frac{d F^{(0)}}
{d \xi} \right|_{\xi=\xi_{p}}
- \left. \frac{d F^{(1)}}{d \xi} \right|_{\xi=\xi_{p}},
\label{adelta2}
\end{eqnarray}
where $F^{(0)}$ and $F^{(1)}$ are solutions of the first and the
second initial-value problem, respectively.

To solve the original boundary-value problem, we have to join
smoothly both solutions, i.e. the conditions
\begin{equation}
\left\{
\begin{array}{rcl}
\Delta (F'_{0},F_{1}) &=& 0 \\
\Delta' (F'_{0},F_{1}) &=& 0
\end{array}
\right. \label{asystem}
\end{equation}
should be satisfied. Eq.(\ref{asystem}) can be considered as
a system of two coupling algebraic equations and can be solved by
the Newton method.

We have found numerically two solutions of Eq.(\ref{asystem}).
Substituting them into Eqs.(\ref{abound1a}) and (\ref{abound1b})
one obtains two pairs of initial condition for the chiral
angle on the interval $[1,\infty)$. Corresponding initial-value
problems can be also solved numerically.

\end{document}